\pdfoutput=1

\documentclass{JINST}
\usepackage[font=small, labelfont=bf]{caption}
\usepackage{subfig}
\usepackage{multirow}
\usepackage{placeins}
\usepackage{textcomp}

\title{Simulation study of a highly efficient, high resolution X-ray sensor based on self-organizing aluminum oxide}

\author{Joerg Muehlbauer$^a$\thanks{Corresponding author.}~, Frank Sukowski$^a$, Nils Reims$^a$, Peter Krueger$^b$, Juergen Schreiber$^b$, Nikolai I. Mukhurov$^c$ and Norman Uhlmann$^a$\\ 
\llap{$^a$}Fraunhofer Development Center X-Ray Technology (EZRT),\\
  Dr.-Mack-Strasse 81, 90762 Fuerth, Germany\\
\llap{$^b$}Fraunhofer Institute for Non-Destructive Testing, Dresden branch IZFP-D,\\
  Maria-Reiche-Strasse 2, 01109 Dresden, Germany\\
\llap{$^c$}Stepanov Institute of Physics of NAS of Belarus,\\
  Ave. Nezalezhnastsi 68, 220072 Minsk, Belarus\\
  E-mail: \email{joerg.muehlbauer@iis.fraunhofer.de}}

\abstract{State of the art X-ray imaging sensors comprise a trade-off between the achievable efficiency and the spatial resolution. To overcome such limitations, the use of structured and scintillator filled aluminum oxide (AlOx) matrices has been investigated. 
We used Monte-Carlo (MC) X-ray simulations to determine the X-ray imaging quality of these AlOx matrices. Important factors which influence the behavior of the matrices are: filling factor (surface ratio between channels and `closed` AlOx), channel diameter, aspect ratio, filling material etc. Therefore we modeled the porous AlOx matrix in several different ways with the MC X-ray simulation tool ROSI \cite{bib1} and evaluated its properties to investigate the achievable performance at different X-ray spectra, with different filling materials (i.e. scintillators) and varying channel height and pixel readout. In this paper we focus on the quantum efficiency, the spatial resolution and image homogeneity.}


\begin{document}

\section{Goals and motivation}
State of the art X-ray imaging sensors comprise a trade-off between the achievable efficiency and the spatial resolution. For instance a thick scintillator yields a high efficiency but suffers from a decrease in spatial resolution due to (optical) light spreading in the scintillation volume. For thinner scintillators the opposite is observed. In general an increase in spatial resolution leads to a decrease in efficiency and vice versa.
The goal of the Fraunhofer-internal project WISA Honeris is to overcome such limitations and to provide `ready for the market' X-ray imaging sensors which offer both high efficiency and high spatial resolution. 
This is achieved by filling structures (hexagonally arranged cylindrical channels) of porous self-organized AlOx matrices with appropriate scintillator materials. Due to its channel-like structures which act as light guides, the high spatial resolution is maintained with increasing scintillator thickness. This permits the fabrication of very thick and therefore highly efficient scintillator matrices without losing spatial resolution.
We will not go into further detail with the fabrication of the AlOx matrices, since this is not the focus of this paper. More information can be found in \cite{bib2}. 
Instead we used Monte-Carlo simulations to investigate the effects of varying matrix parameters and different scintillation materials regarding spatial resolution and quantum efficiency. The goal of this simulation study is to generate a priori knowledge about the behavior of the scintillator matrix  and to examine various questions which are difficult to answer in real experiments. This should help in the fabrication and optimization of the matrices.

\section{Monte-Carlo model}
\begin{figure}[htbp]
  \centering
	\centering
    \includegraphics[width=.3\textwidth]{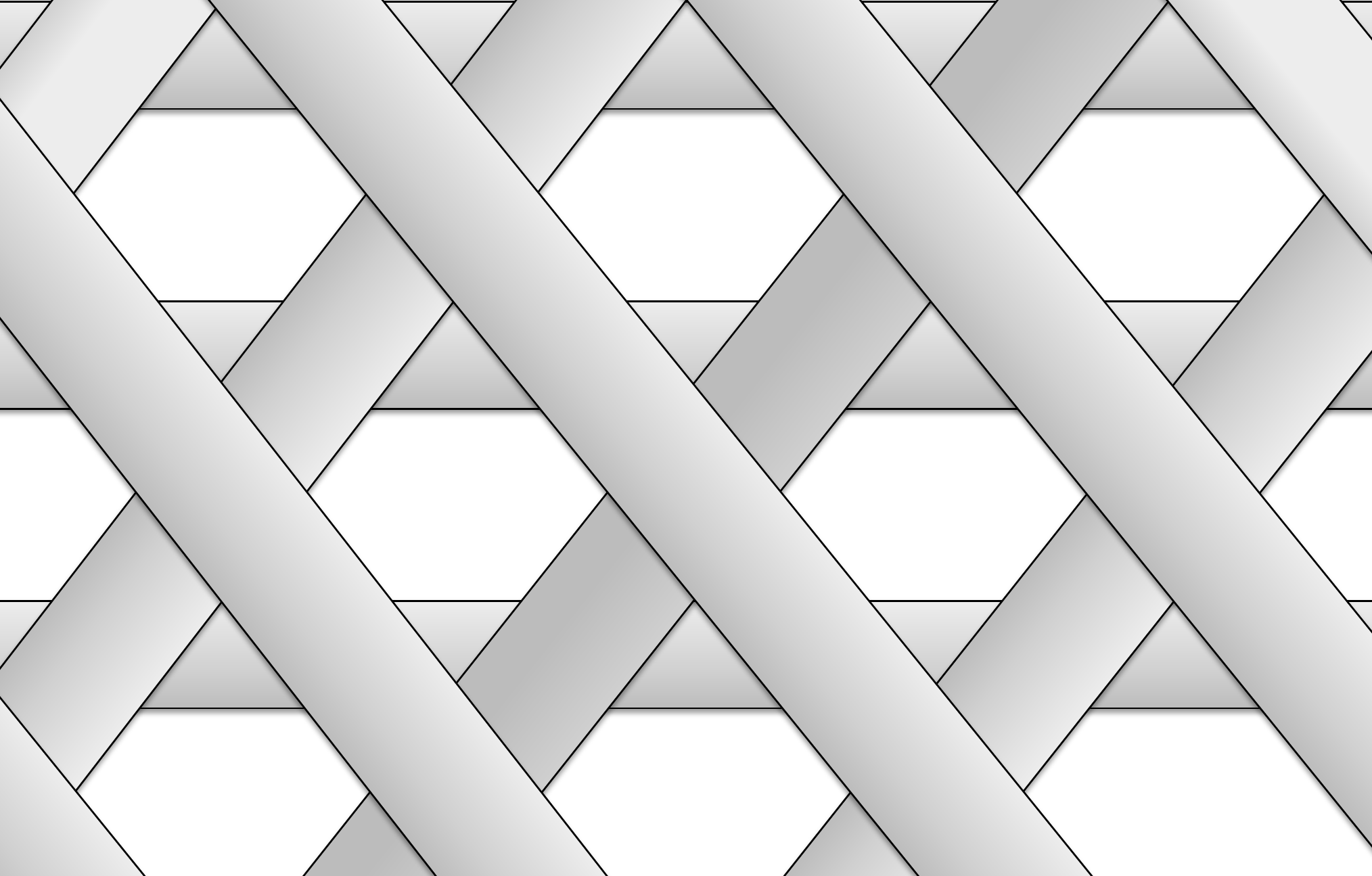}  
    \caption{Simplified simulation model.}
    \label{fig:hex_structure_thick}
\end{figure}

We constructed a model of the matrix structure in ROSI with adjustable parameters like channel diameter $d_c$, channel height $h_c$, channel pitch (center to center) $p_c$ and scintillation material. For the following simulations a $d_c$ of \mbox{1.6 \textmu m}  and a  $p_c$ of 2 \textmu m were used. 
To reduce the simulation time we approximated the matrix structure by a simplified model consisting of a massive scintillator medium divided by diagonal and horizontal strips of certain thickness (see figure \ref{fig:hex_structure_thick}). The filling factor was about 36 \%. The irradiation was simulated with a tungsten X-ray tube at varying tube voltages $U_t$.

\section{Quantum Efficiency} \label{sec:qe}
\begin{figure}
	\centering
		\includegraphics[width=0.75\textwidth]{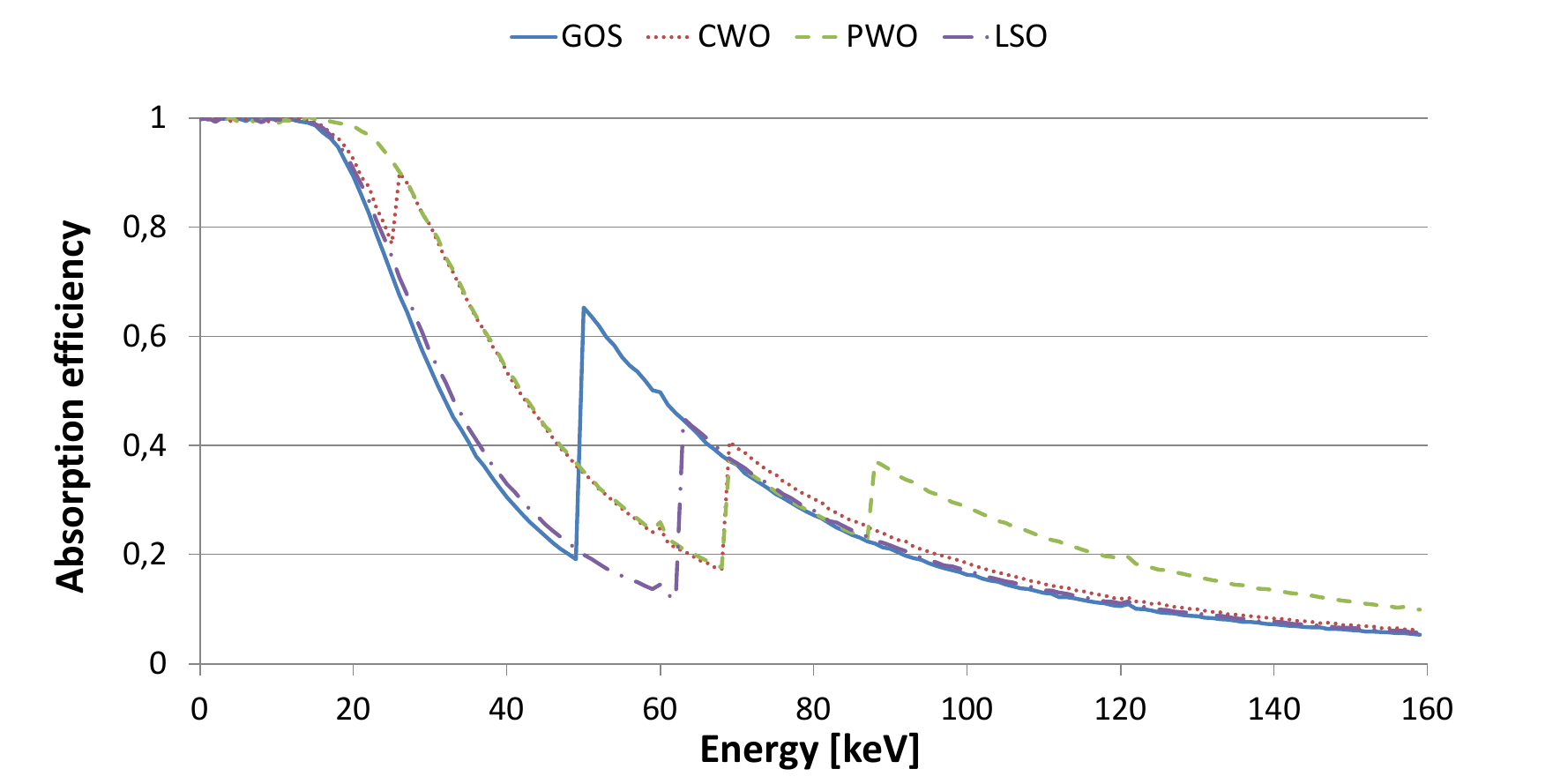}
	\caption{Absorption curve for selected materials with thickness 100 \textmu m.}
	\label{fig:absorption}
\end{figure}

The simulations regarding quantum efficiency (QE) were carried out with different scintillation materials, namely $\rm{Gd_2O_2S}$ (GOS), $\rm{CdWO_4}$ (CWO), $\rm{PbWO_4}$ (PWO) and $\rm{Lu_2SiO_5}$ (LSO). To determine the absorption properties of the matrix structure consisting of the scintillator channels and the surrounding AlOx volume, an area of $4\times4$ pixels with a size of 4 \textmu $\rm{m}^2$ each  and $h_c$ of \mbox{100 \textmu m} and \mbox{500 \textmu m} was irradiated with both $U_t=35$ kV and $U_t=160$ kV. Based on the deposited energy in the scintillator channels, we calculated the respective efficiency (excluding loss due to optical effects and limited readout filling factor, which were not simulated).
The absorption probability curve for the homogeneous materials with $h_c=100$ \textmu m is shown in \mbox{figure \ref{fig:absorption}}. Although the heavy materials PWO and CWO show a better mean probability over the whole energy range, GOS still reaches a good overall efficiency due to its (Gadolinium) K-edge at about \mbox{50 keV}, which is close to the K-$\alpha$ emission line of the applied tungsten X-ray spectra. 
The absorption properties of the whole scintillator matrix are shown in table \ref{tab:tab1}. As expected, the \mbox{$U_t=35$ kV} case showed nearly similar results for all tested scintillator materials. With a QE near 100 \% for all scintillators at the center of mass energy of the 35 kV spectrum at about 15 keV (see figure \ref{fig:absorption}), the efficiency is mostly determined by the AlOx volume which is identical in all cases. The deviation between the materials is below 4 \%. Simulations with $U_t=35$ kV at $h_c=500$ \textmu m were skipped because of similar results to be expected. With $U_t=160$ kV the differences were more significant. PWO performed best with \mbox{7.8 \%} \mbox{(23.9 \%)} efficiency while GOS reached \mbox{6.9 \%} \mbox{(20.2 \%)} for $h_c=100$ \textmu m (500 \textmu m). The results for CWO were quite similar with 6.9 \% (22.6 \%). LSO showed the worst QE with only 5.2 \% (18 \%). 

\begin{table}[!h]
\centering
\begin{tabular}{c|cccccccc}
		&	\multicolumn{2}{c}{GOS}	& \multicolumn{2}{c}{CWO}	& \multicolumn{2}{c}{PWO} & \multicolumn{2}{c}{LSO} \\
		& 100 \textmu m	&	500 \textmu m	& 100 \textmu m	&	500 \textmu m &	 100	\textmu m &	500	 \textmu m & 100	\textmu m &	500 \textmu m \\
\hline
$U_t=35$ kV & 47 & - & 47.9 & - & 49.6 & - & 47.2 & - \\
$U_t=160$ kV & 6.9 & 20.2 & 6.9 & 22.6 & 7.8 & 23.9 & 5.2 & 18 \\ 
\end{tabular}
\caption{Quantum efficiency for the AlOx matrix filled with different scintillator materials, as a function of $h_c$ and $U_t$. The simulations for $h_c=500$ \textmu m at $U_t=35$ kV were skipped because of the negligible effect of the scintillator thickness on the efficiency at these low energies. The numbers are given in percent.}
\label{tab:tab1}
\end{table}
\FloatBarrier

\section{Image homogeneity}
\begin{figure}[htbp]
  \centering
    \begin{minipage}[t]{7 cm}
	\centering
    \includegraphics[width=0.9\textwidth]{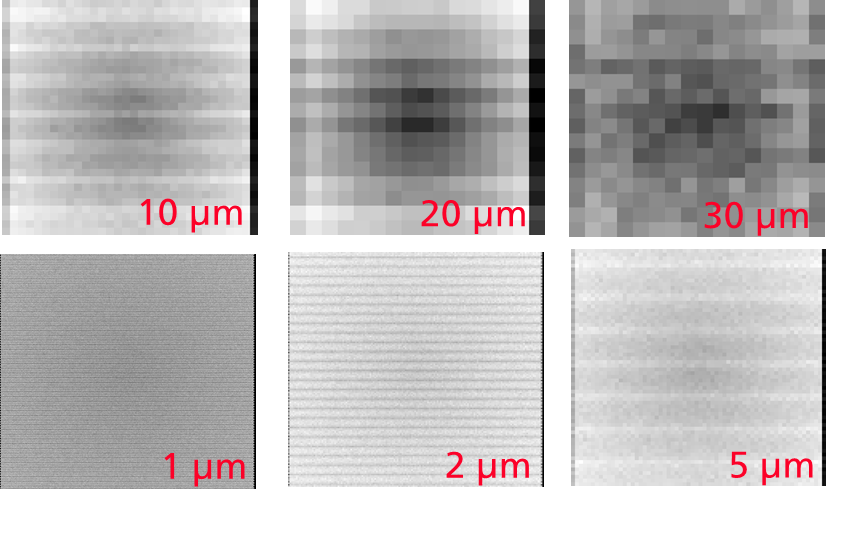}  
   \caption{Intensity modulation and inhomogeneities with different pixel sizes.}
    \label{fig:intensity_mod}
    \end{minipage}
    \quad
  \begin{minipage}[t]{7 cm}
	\centering
    \includegraphics[width=0.9\textwidth]{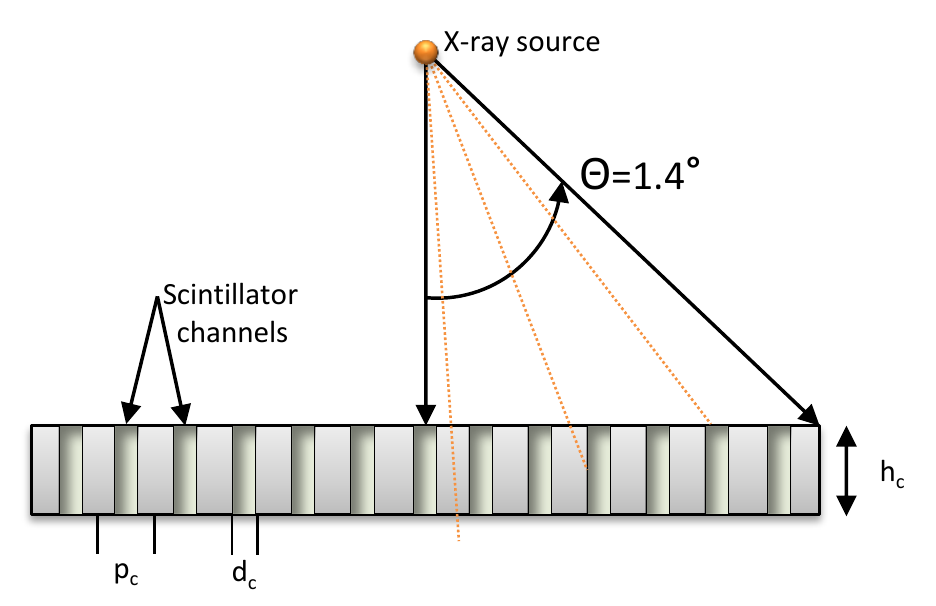} 
    \caption{Irradiation setup for the simulations regarding spatial resolution. Not true to scale. }
	\label{fig:angle}
  \end{minipage}
\end{figure}

To examine the influence of the number of channels per readout pixel on the image homogeneity, the hexagonal structure was read out by an orthogonal pixel matrix with varying pixel sizes ranging from 1 \textmu m to 30 \textmu m. The scintillator matrix was irradiated with $U_t=40$ kV.
The superposition of two periodical grids, in this simulation setup, the scintillator matrix and the pixelated readout (sensor) leads to an intensity modulation as seen in figure \ref{fig:intensity_mod}. The modulation is about 15 \%. This is due to aliasing errors (Moiré-pattern). The frequency of the modulation varies with the number of channels per readout pixel. The more channels per pixel the lower the modulation frequency. 
One phenomenon, which was observed during these simulations, was an intensity gradient from the center towards the outside of the scintillator matrix. This effect becomes more significant with an increasing number of channels per readout pixel. 
To take a closer look at this effect, the scintillator matrix was simulated with a fixed readout pixel matrix of 20 \textmu m pixel pitch and $h_c$ of 100 \textmu m and 500 \textmu m, irradiated with $U_t$ of 40, 80, 120 and 160 kV. We used a virtual Focus-Detector-Distance (FDD) of 100 cm and irradiated one half of the scintillator matrix of 4.8 cm edge length with a cone beam. This results in X-rays impinging the matrix at an angle $\theta$ between $0^\circ$ and $1.4^\circ$ with respect to the normal of the matrix surface (see figure \ref{fig:angle}).

\begin{figure}[htbp]
  \centering
  \begin{minipage}[b]{7 cm}
    \includegraphics[width=0.95\textwidth]{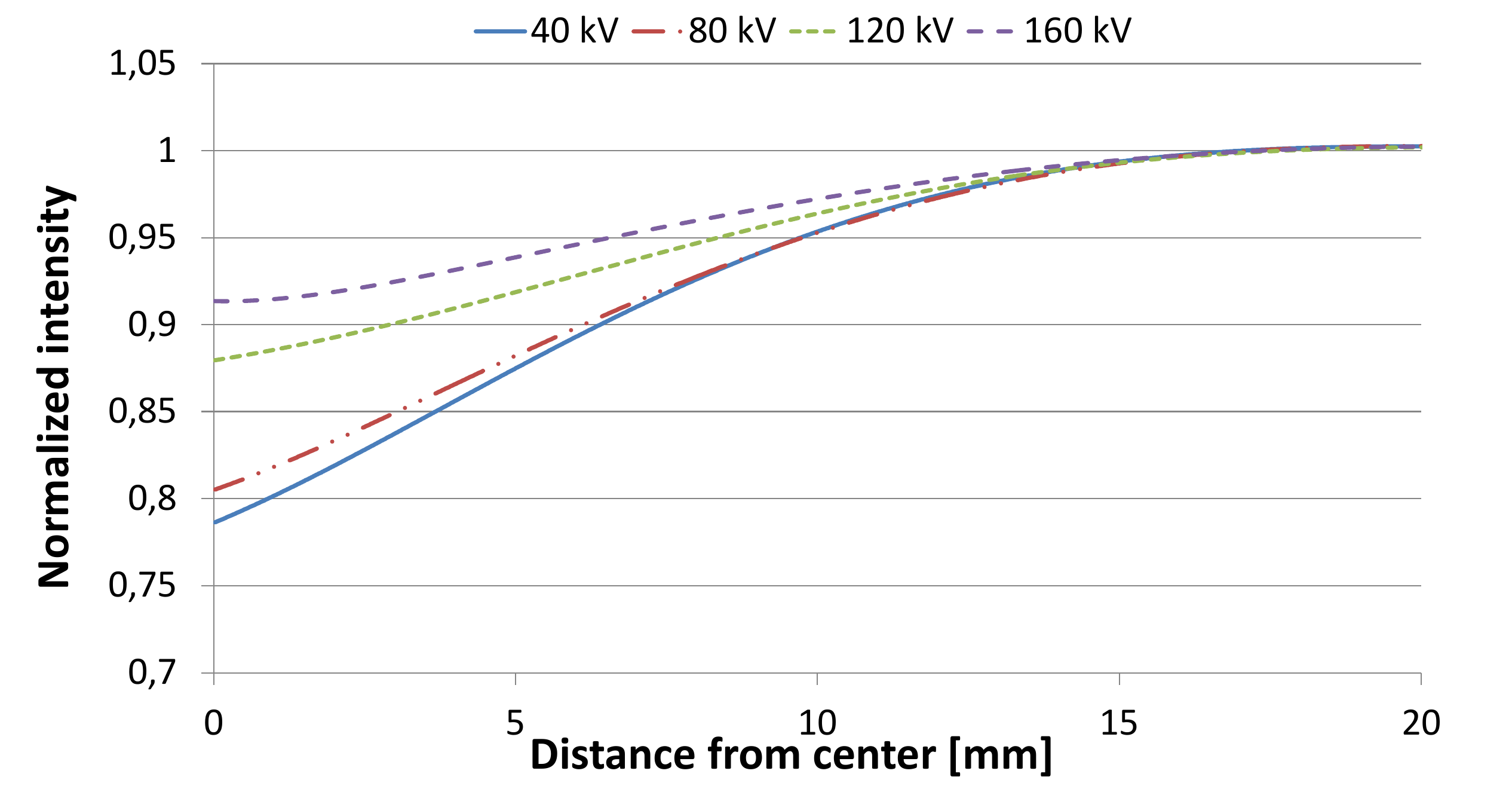}  
    \caption{Intensity gradient for \mbox{$h_c= 100$ \textmu m.}}
    \label{fig:intensity_mod_100}
    \end{minipage}
    \qquad
  \begin{minipage}[b]{7 cm}
    \includegraphics[width=0.95\textwidth]{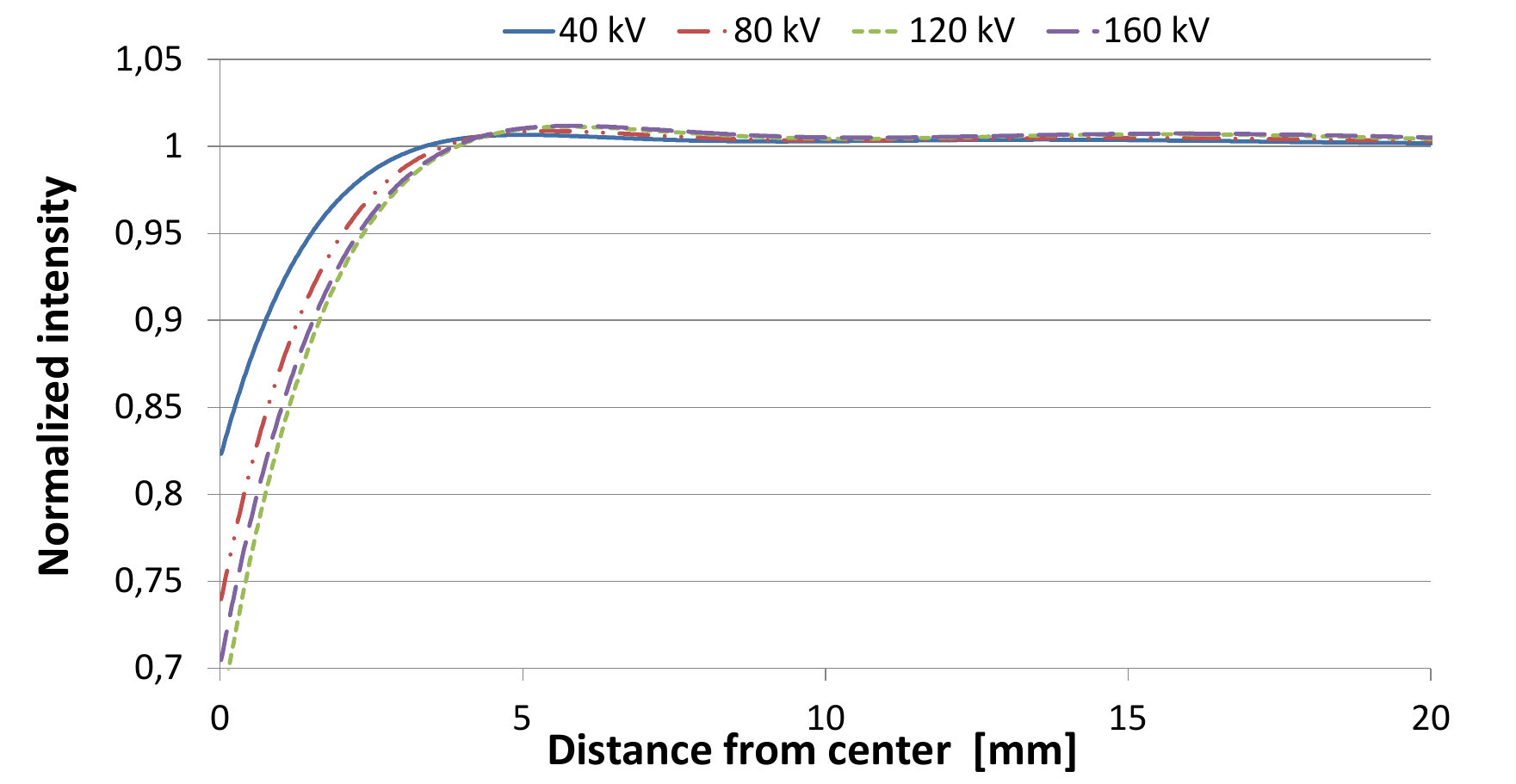} 
    \caption{Intensity gradient for \mbox{$h_c=500$ \textmu m.}}
    \label{fig:intensity_mod_500}
  \end{minipage}
\end{figure}

Unlike a homogeneous scintillator medium where the intensity decreases towards the edges of the medium, a significant drop of intensity can be seen in the center of the simulated matrix. This intensity gradient depends on $U_t$ and $h_c$ as can be seen in figure \ref{fig:intensity_mod_100} and \ref{fig:intensity_mod_500}.
The gradient results from X-ray photons impinging the matrix at an angle. More precisely, it depends on the integral
 scintillator thickness, which an X-ray photon penetrates on its trajectory through the matrix until absorption or leaving of the structure. In the simulated setup (fixed channel diameter and fixed pixel pitch) this in turn depends on $h_c$ and the X-ray photon energy.
With $h_c=100$ \textmu m and a low mean photon energy like with $U_t=40$ kV, the mean penetration length is rather small, which means the absorption happens mainly at the surface of the scintillator channel. For simplification one could think of the surrounding AlOx material as zero absorbing and the scintillation channels as fully absorbing. Then photons at a very small angle have a higher chance to pass the scintillator matrix without hitting a scintillator channel (see figure \ref{fig:angle}). This results in a significant drop of absorbed photons, i.e. less intensity in the area near the center. Increasing $h_c$ reduces this angle where no absorption happens. This results in a steeper curve as can be seen in figure \ref{fig:intensity_mod_500}. With higher photon energies the assumption that absorption mainly takes place at the surfaces is no longer true. 
The probability to be transmitted through scintillator channels increases. Furthermore photons are more likely to interact in adjacent channels if the incident angle is high enough.
The integral scintillator thickness gets more important. For $h_c=100$ \textmu m this effect results in a less significant drop of intensity near the center with increasing energy. The $h_c=500$ \textmu m case shows an inverse effect. This means there must be another effect which affects the gradient. Although not investigated in detail we assume, that this has to do with beam hardening. The AlOx parts with $h_c=500$ \textmu m can no longer be seen as zero absorbing.

Imagine two photons $p_1$ and $p_2$. $p_1$ impinges the scintillator matrix at such a small angle that it will go through the matrix without hitting a scintillator channel on its trajectory for both $h_c=100$ \textmu m and 500 \textmu m. On the contrary $p_2$ impinges the matrix at a slightly larger angle, so that it only hits a scintillator channel if $h_c$ is at least 500 \textmu m. In the 100 \textmu m case the energy $E$ of the particles $p_1$,  $p_2$ is not important. They either get absorbed in the AlOx volume or are transmitted through the matrix. In both cases no signal is generated. The 500 \textmu m case however shows an energy dependence for $p_2$. For a small $E_{p2}$, $p_2$ is more likely to get absorbed in the AlOx volume because of the long distance to reach the scintillator channel. In contrary if $E_{p2}$ is high, $p_2$ will probably reach the channel and deposit some energy. That means the difference between $p_1$ and $p_2$ increases with increasing photon energy as can be seen in figure \ref{fig:intensity_mod_500}. An X-ray spectrum with a higher mean photon energy results in a steeper curve in the small angle area around the center.

\section{Basic spatial resolution}
\begin{figure}[htbp]
  \centering
  \subfloat[\label{fig:bsr_0_35kV}]{\includegraphics[width=.48\textwidth]{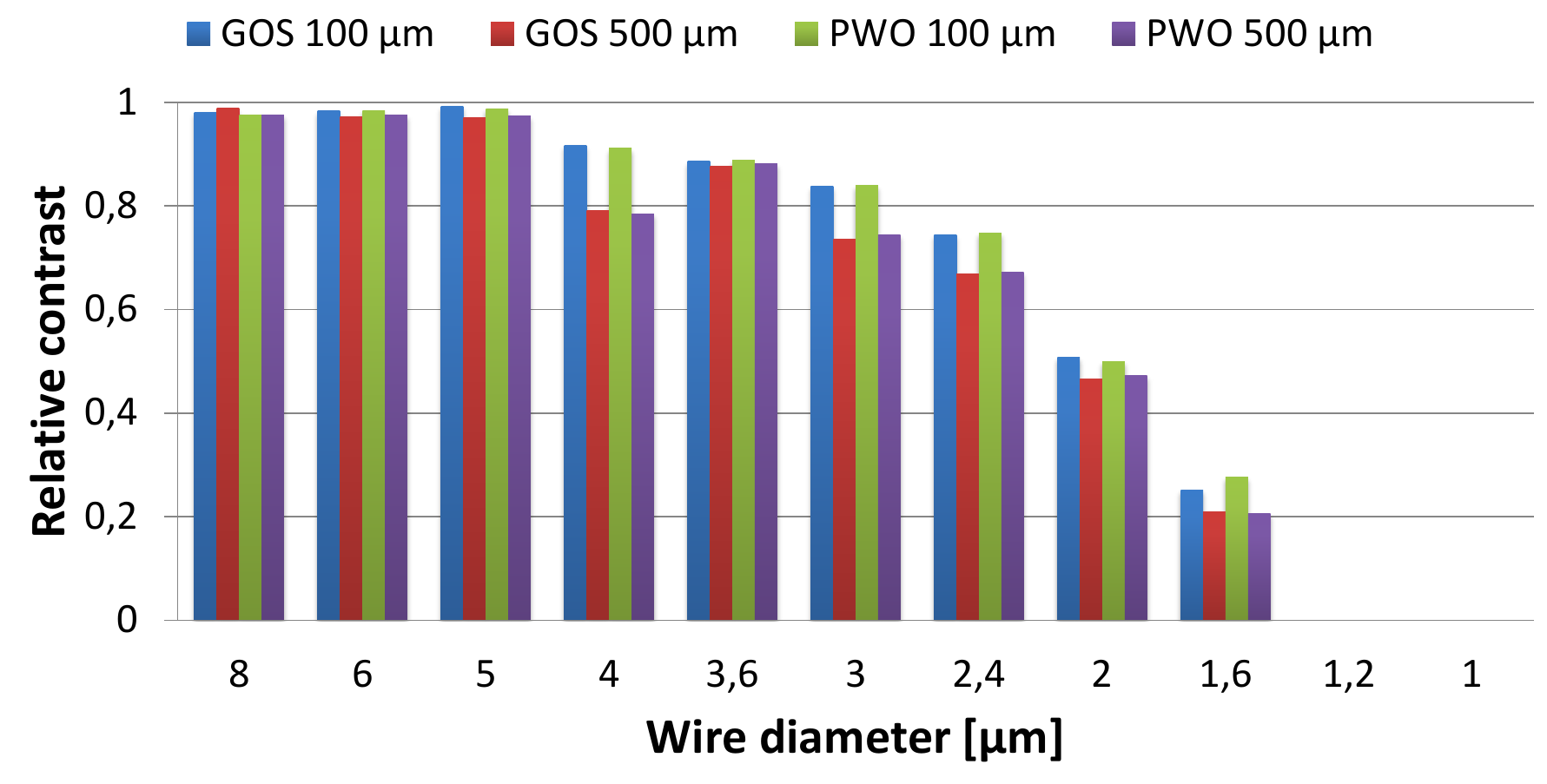}}  
    \subfloat[\label{fig:bsr_0_160kV}]{\includegraphics[width=.48\textwidth]{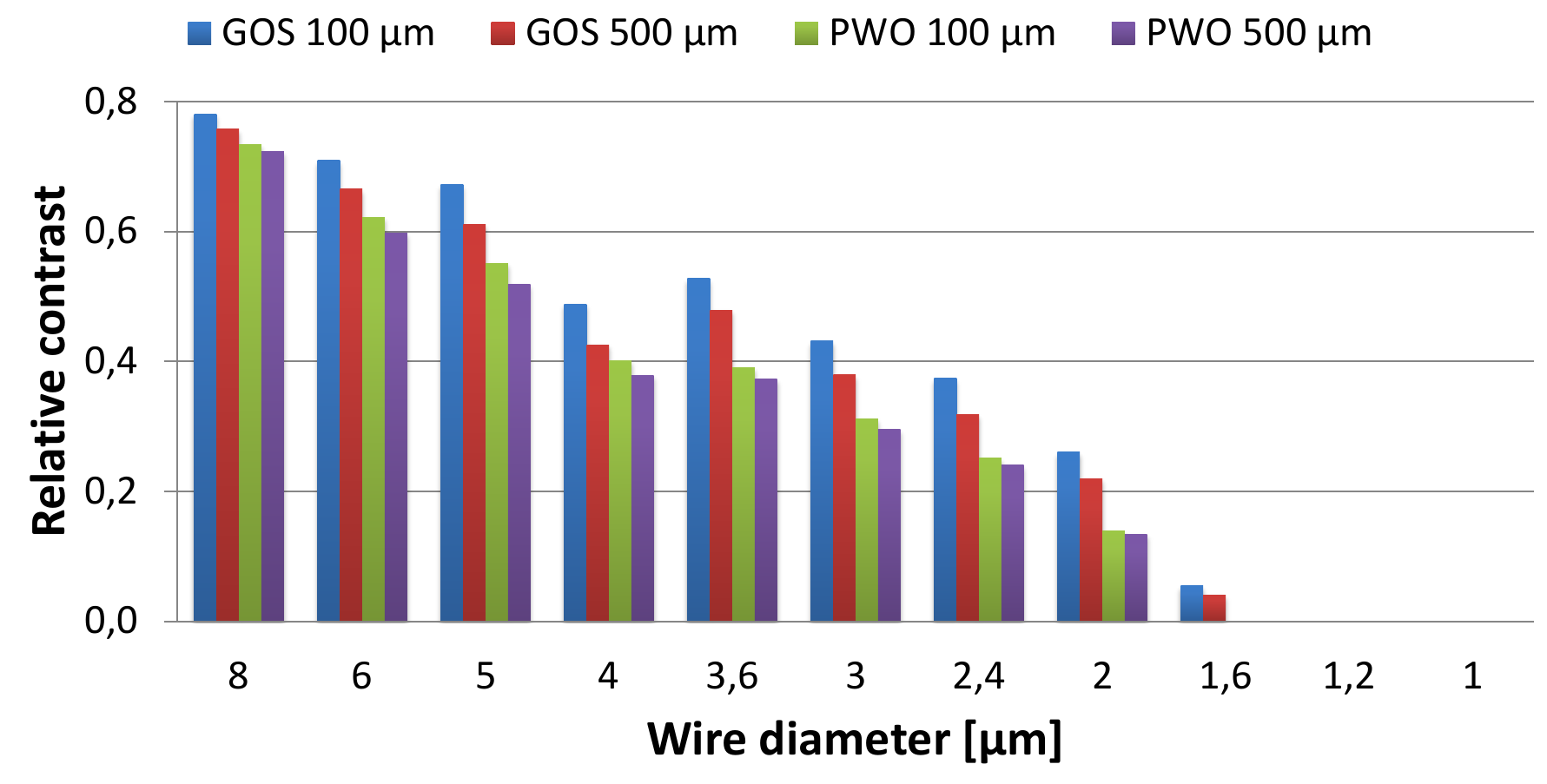} }\\
    \subfloat[\label{fig:bsr_1p4_35kV}]{\includegraphics[width=.48\textwidth]{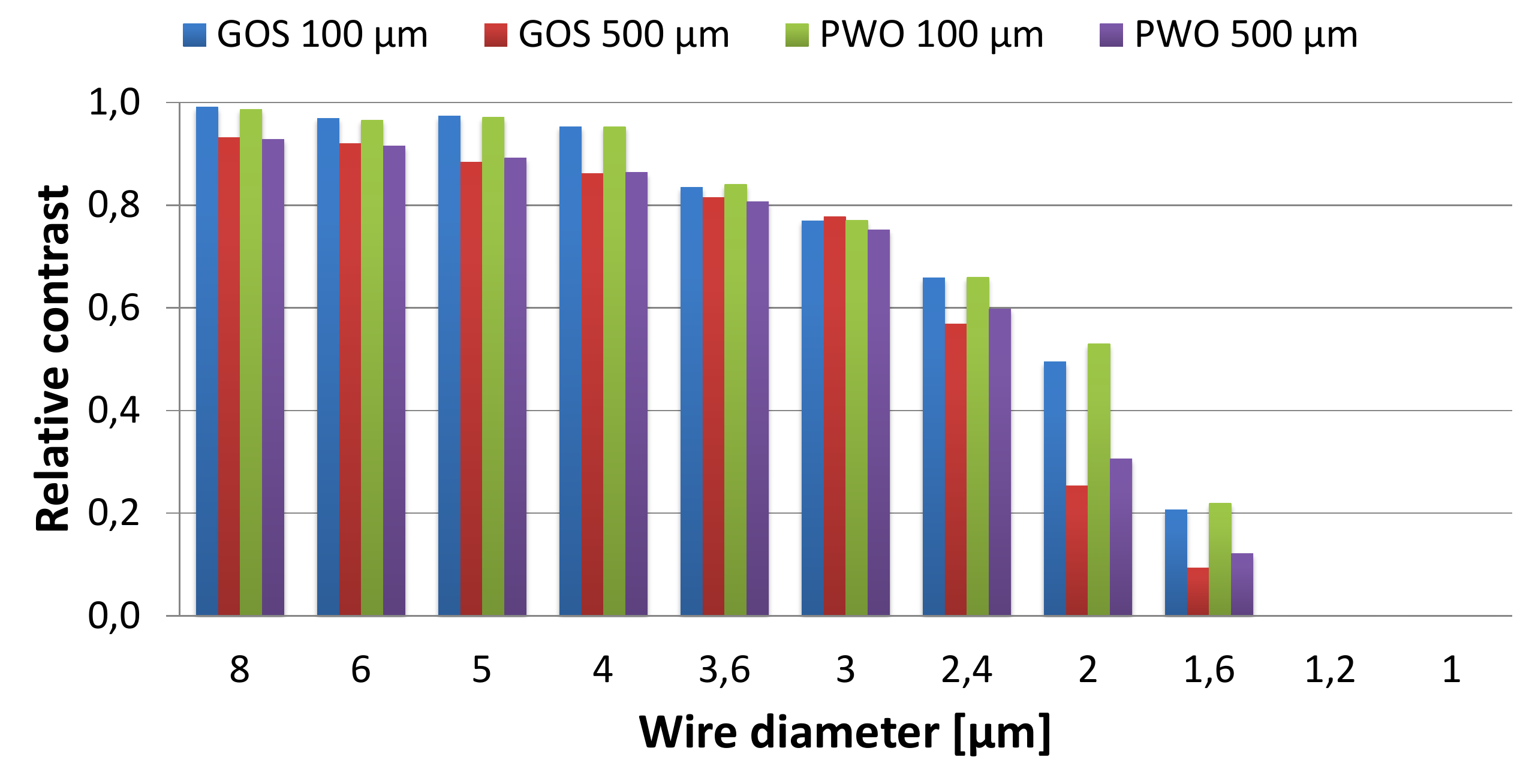}}
	\subfloat[\label{fig:bsr_1p4_160kV}]{\includegraphics[width=.48\textwidth]{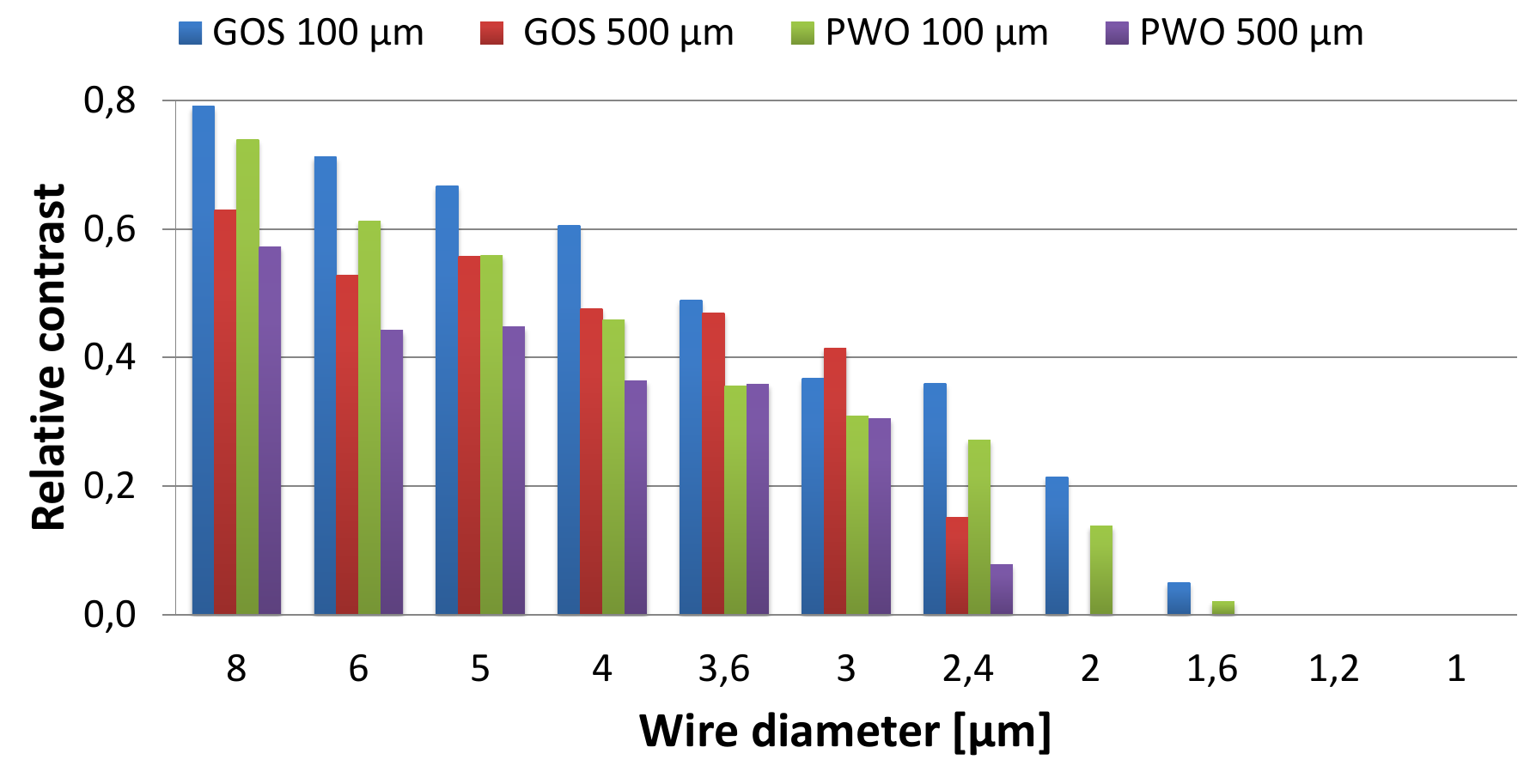}}  
	\caption{Image contrast based on BSR method for GOS and PWO with \textbf{(a)} parallel beam and $U_t = 35$ kV, \textbf{(b)} parallel beam and $U_t = 160$ kV, \textbf{(c)} cone beam and $U_t = 35$ kV, \textbf{(d)} cone beam and $U_t = 160$ kV. The local minima at the 4 \textmu m wire pair in figure a and b result from aliasing errors caused by the pixel readout.}
	\label{fig:contrast_bsr}
\end{figure}

We used the basic spatial resolution (BSR) method to study the spatial resolution as a function of the scintillation material, $U_t$ and $h_c$. It is part of a well established detector characterization standard \cite{bib3}. Compared to an MTF calculation, the BSR method is a more practical approach for measuring the visual spatial resolution. It utilizes a plate of acrylic glass containing pairwise arranged platinum and tungsten wires of decreasing diameter. It is placed directly in front of a detector. From the gray value profile perpendicular to the wire length, one can calculate the contrast of the wires in relation to the area between the wires. The BSR value is defined as the (interpolated) wire diameter with exactly 20 \% contrast. Considering the very high spatial resolution to be expected, we modeled  a modified double wire phantom with thinner wire diameters than specified in the standard, ranging from 8 \textmu m to 1 \textmu m. The simulated scintillation materials were GOS with the highest light yield  of all considered materials \cite{bib4}, \cite{bib5} and PWO for the highest QE (see table \ref{tab:tab1}). Two different irradiation setups were taken into account. In addition to the cone beam irradiation with $\theta$ between $-1.4^\circ$ and $+1.4^\circ$ we also investigated a parallel beam geometry, where every wire pair is irradiated under the same angle $\theta = 0^\circ$. The cone beam setup corresponds to a realistic setup with an FDD of 100 cm and a scintillator matrix edge length of 4.8 cm. Each case was simulated with \mbox{$U_t=35$ kV}, \mbox{160 kV} and with \mbox{$h_c=100$ \textmu m}, \mbox{500 \textmu m}. 
The simulated image contrast for each wire pair can be seen in figure \ref{fig:contrast_bsr}, the corresponding calculated BSR values in table \ref{tab:bsr_values}.
With an X-ray spectrum of $U_t=35$ kV, corresponding to a small penetration length, we observed only small differences between the two scintillators. At parallel beam irradiation both materials show nearly the same spatial resolution with both $h_c=100$ \textmu m and 500 \textmu m. However, for cone beam irradiation the spatial resolution is significantly degraded (up to 17 \% compared to the parallel case) by increasing $h_c$ from 100 \textmu m to 500 \textmu m because of the higher probability of X-rays impinging the matrix structure at an angle, to hit a scintillator channel. The difference between both materials is  below 6 \% due to the low Compton scattering cross section in this X-ray energy range.

An X-ray spectrum of $U_t=160$ kV, corresponding to a higher penetration length, leads to an overall lower spatial resolution than the 35 kV case because of the generally higher Compton scattering cross section. Furthermore the higher X-ray energy increases the differences between the two tested materials (up to 9 \%). At parallel beam irradiation only a small degradation can be seen when increasing $h_c$ from 100 \textmu m to 500 \textmu m. At cone beam irradiation the spatial resolution depends on both material and channel height. \\
The material dependency is caused by the different X-ray interaction cross sections. As can be seen from figure \ref{fig:contrast_bsr}, at $U_t=35$ kV PWO reaches slightly better contrasts than GOS due to its higher absorption capability while at $U_t=160$ kV GOS clearly surpasses PWO. We assume that this is directly related to the high energy X-ray fluorescences of Pb (75 keV), which occur at energies above 88.3 keV and are isotropically emitted. Compared to the Gd fluorescences of \mbox{43 keV}, they have a much higher penetration length and therefore a higher impact on the degradation of the spatial resolution.\\
The $h_c$ dependency is caused by the higher probability to absorb a scattered X-ray photon for higher channels.

\begin{table}
\centering
\begin{tabular}{ccc|ccccc}
Irradiation	&	$h_c$	& $U_t$	 &  $\rm{BSR_{GOS}}$	&	$\rm{BSR_{PWO}}$ & [\textmu m]\\	
\hline
\multirow{4}{*}{Parallel beam} & 	\multirow{2}{*}{100 \textmu m} & 35 kV & 1.6 & 1.6 &\\
	&		&160 kV & 1.9 & 2.2 & \\
	&	\multirow{2}{*}{500 \textmu m} & 35 kV & 1.6 & 1.5 & \\
	&		&160 kV & 2.0 & 2.2 & \\	
\hline
\multirow{4}{*}{Cone beam} & 	\multirow{2}{*}{100 \textmu m} & 35 kV & 1.6 & 1.6 & \\
	&		&160 kV & 2.0 & 2.2 & \\
	&	\multirow{2}{*}{500 \textmu m} & 35 kV & 1.9 & 1.8 & \\
	&		&160 kV & 2.5 & 2.7 & \\	
\end{tabular}
\caption{BSR values for the scintillator materials GOS and PWO, as a function of irradiation setup, $h_c$ and $U_t$. The numbers are given in \textmu m.}
\label{tab:bsr_values}
\end{table}

\section{Conclusion and outlook}
The goal of this simulation study was to get insight in the behavior of structured scintillators without the need to fabricate multiple test samples. Therefore we used a simplified simulation model and studied the efficiency and spatial resolution with different materials and geometries.
The simulations clearly showed the achievable performance of this technique but also gave insight in the forthcoming problems and potential solutions. GOS turned out to be the most suitable of the tested materials due to its high spatial resolution, light yield and decent QE. Heavy materials like PWO could not compensate their far inferior light yield with their high QE which is mainly determined through the dominating AlOx parts. In addition they did not show a better spatial resolution though their lower scattering contribution as shown in the BSR simulations. 
\newline
Further research has to be done on the topic of aliasing errors, which are to be expected because of the special channel-like geometry including methods to minimize these. 
In the meantime first prototypes based on the simulation results have been produced and are experimentally investigated and compared to the simulations. This will be subject to an upcoming publication.

\acknowledgments
This work was supported by the FhG Internal Programs under Grant No. WISA 820 229

\end{document}